\documentclass[12pt]{iopart}

\usepackage{graphicx}
\usepackage{iopams}

\def\der#1#2{{\partial#1 \over \partial#2}}
\def\be{\begin{equation}}
\def\ee{\end{equation}}
\def\ba#1{\begin{array}{#1}}
\def\sx#1{\hat{\sigma}^x_{#1}}

\def\sz#1{\hat{\sigma}^z_{#1}}
\def\ea{\end{array}}
\def\bn{\begin{enumerate}}
\def\en{\end{enumerate}}

\def\rr{\right}
\def\l{\left}

\def\H{{\cal H}}
\def\summ{\sum\limits}

\def\tr{{\mathrm tr}}

\def\G{\Gamma}
\def\ket#1{\l|#1\rr\rangle}
\def\bra#1{\l\langle#1\rr|}
\def\intt{\int\limits}
\def\summ{\sum\limits}

\def\beq{\begin{equation}}
\def\eeq{\end{equation}}
\def\S{\overline{S}}

\begin{document}

\title{Criticality and entanglement in random quantum systems}
\author{G. Refael}
\address{Department of Physics, California Institute of
  Technology, MC 114-36,1200 E. California Blvd.,  Pasadena, CA 91125}
\author{J.~E.~Moore}
\address{Department of Physics, University of California,
Berkeley, CA 94720}
\address{Materials Sciences Division,
Lawrence Berkeley National Laboratory, Berkeley, CA 94720}
\begin{abstract}
We review studies of entanglement entropy in systems with quenched
randomness, concentrating on universal behavior at strongly random
quantum critical points.  The disorder-averaged entanglement entropy
provides insight into the quantum criticality of these systems and an
understanding of their relationship to non-random (``pure'') quantum
criticality.  The entanglement near many such critical points
in one dimension shows a logarithmic divergence in subsystem size,
similar to that in the pure case but with a different universal
coefficient.  Such universal coefficients are examples of universal
critical amplitudes in a random system.  Possible measurements are
reviewed along with the one-particle entanglement scaling at certain Anderson
localization transitions. We also comment briefly on higher dimensions
and challenges for the future.

\end{abstract}

\maketitle

\section{Introduction}


Several important connections have been established in recent years between concepts from quantum information and problems in many-body physics.  One such connection is the use of entanglement entropy to understand how ground states of various quantum Hamiltonians show either criticality or topological order.  This article reviews recent work on entanglement entropy in systems with quenched randomness, and will focus for the most part on universal behavior connected with strongly random quantum critical points.  Beyond improving the general understanding of entanglement in many-particle systems, these studies have given useful insight into random quantum critical points.  For example, entanglement entropy provided the first example of a universal ``critical amplitude'' at this type of critical point.

In this introduction, we review the basic concept of entanglement entropy and a few of its applications to non-random systems, then introduce the physics of strongly random quantum critical points.  These critical points are best understood in one spatial dimension, where the real-space renormalization group (RSRG) approach gives nonperturbative results for many disordered-averaged quantities that are believed to be exact.  In some cases, such as the random XX spin chain that is connected to free fermions with random hopping via the Jordan-Wigner transformation, numerical results can confirm the RSRG predictions.  We then discuss the possibility of experimental measurements and make some introductory comments about strongly random systems in higher dimensions.  The later sections of this review are as follows: Sections 2 and 3 introduce infinite-randomness fixed points and calculations of their entanglement, Section 4 discusses possible experimental observations, Section 5 reviews single-particle entanglement in Anderson-type localization problems, Section 6 discusses numerical results, and Section 7 briefly introduces higher-dimensional results and some open problems for the future.

\subsection{Entanglement entropy}


A fundamental concept of quantum information is the
entanglement entropy of a pure quantum state, defined as the von Neumann entropy
of the reduced density matrix created by a partition of the system
into parts $A$ and $B$:
\beq
S = - {\rm Tr}\ \rho_A \log_2 \rho_A = - {\rm Tr}\ \rho_B \log_2 \rho_B.
\eeq
Here the base-2 logarithm means that the entropy is measured in bits.  If the original state is not a product state (i.e., does not factorize into a pure state for subsystem $A$ and one for $B$), then the entanglement entropy is nonzero.  Since it is determined by the reduced density matrix for a subsystem, which characterizes all physical measurements on that subsystem, there is no way to distinguish through measurements only on $A$ whether an entropy arose from partition of an entangled pure state of $AB$ or from a mixed state of $AB$.

In most cases we will be interested in partitions of an infinite
system into a finite contiguous part $A$ and a remainder $B$.
Entanglement of the ground state of a local Hamiltonians is different
from that of a generic state in the Hilbert space: the locality means
that ground states with an energy gap to excitations typically show
entanglement that scales with the size of the $AB$ boundary (the
``area law''~\cite{Srednicki93,Hastings:2007p1181,Eisert:2008p1171}).
This leads to the notion of a length scale around the boundary beyond
which entanglement must decay rapidly.  As reviewed in the following
section, quantum critical points can violate the area law because,
just as correlations become long-ranged at a quantum critical point,
entanglement can do so as well.  The increase at entanglement at
quantum critical points seems to be a general property of both random
and translational-invariant systems, and is clearest in one dimension
as explained below.

A question occasionally raised is to what extent entanglement entropy and other quantum information concepts are either ``useful'' or ``measurable'' in condensed matter systems.  The first use of quantum information, as in the above examples is to understand something new about condensed matter systems that is not evident in correlation functions or other more conventional quantities.  Another important use, which we will not discuss here, arises from the fact that the entanglement entropy of a state is related to the accuracy with which it can be approximated by the matrix product states that are convenient for numerical simulations.  Two examples from one-dimensional translationally invariant systems are that states satisfying the ``area law'' are described exponentially well by matrix product states, while for critical states~\cite{Tagliacozzo:2008p1188,Pollmann:2008p1189} the matrix product state description converges at a slower rate determined by the critical point's central charge.  Finally, while determination of the entanglement entropy by direct measurement of the reduced density matrix~\cite{Moehring:2007p1283} indeed seems impracticable on a many-particle system, the variance of flux measurements on a spin chain~\cite{KlichRefaelSilva} can be used to obtain the entanglement entropy.  This proposed measurement is closely connected to efforts to determine the source of unusual flux noise in SQUID systems.

Two ways to generalize the entanglement entropy are via the ``Renyi entropy'' of the reduced density matrix
\beq
S_\alpha = {1 \over 1 - \alpha} \log_2 {\rm Tr} {\rho_A}^\alpha,
\eeq
which reduces to the (von Neumann) entanglement entropy as $\alpha \rightarrow 1$,
or the ``entanglement spectrum'' (the full eigenvalue distribution of the reduced density matrix).  The entanglement spectrum has been analyzed for phases with topological order~\cite{Li:2008p1185} or translation-invariant quantum criticality~\cite{Calabrese:2008p1439,Pollmann:2008p1189}, and is discussed briefly below for the random singlet phase.  At least for the random singlet phase, the entanglement spectrum and Renyi entropy are likely to follow quite directly from the ordinary entanglement entropy, but for more complicated random critical points this is not expected to be the case.

\subsection{Quantum criticality}


The entanglement entropy of the ground state at a quantum
critical point can, in some cases, be understood via the
quantum-to-classical mapping.  An important example is furnished by the quantum
critical points in one dimension that become two-dimensional conformal
field theories (CFTs), where the entanglement entropy in the quantum
theory has a logarithmic divergence, whose coefficient is connected to the central charge of the
CFT:~ \cite{Holzhey94,Vidal03,Calabrese04,Ryu1}
\beq
\lim_{N \rightarrow \infty} S \sim {c \over 3} \log_2 N,
\eeq
where we consider $A$ as a finite contiguous set of $N$ spins (or other local Hilbert space) and $B$ is the complement of $A$ in the
infinite chain.  Away from criticality, the entanglement $S$ is
bounded above as $N \rightarrow \infty$ (the one-dimensional version
of the ``area law''.\cite{Srednicki93})
Surprisingly, the entanglement entropy, whose definition is closely tied to the
lattice via the Hilbert space is actually a universal property of the critical field theory,
and hence independent of lattice details.  Most one-dimensional translation-invariant critical points fall into this class: the key is being able to rescale space and time to obtain rotational invariance and thence conformal invariance (for example, $z=2$ quantum critical points do not fall into this class).  At this time we lack a
similarly complete understanding of translation-invariant critical points in higher
dimensions; isolated solvable cases include free
fermions~\cite{gioev,wolf}, higher dimensional conformal field
theories~\cite{Ryu1}, and one class of $z=2$ quantum critical
points.\cite{fradkinmoore}


The connection between the central charge of CFT's and their
entanglement entropy implies that indeed for quantum critical
points with classical analogs, the natural measure of universal
critical entropy in the quantum system (the entanglement entropy) is
determined by the standard measure of critical entropy in the
classical system (the central charge). In addition, it translates
important notions about the central charge to the realm of the
universal quantum measure - the entanglement entropy. Zamolodchikov's
$c$-theorem \cite{zamolodchikov} states that the central charge $c$
decreases along unitary renormalization-group (RG) flows. Therefore we conclude
that the entanglement entropy of CFT's also decreases along RG
flows. Stated this way, the strength of the $c$-theorem may apply to universal critical entropies in quantum systems that are
not tractable by the quantum-to-classical mapping.

Understanding universal behavior near quantum critical points has been
a major goal of condensed matter physics for at least thirty years.
Quantum critical points describe continuous phase transitions at zero
temperature, where quantum-mechanical phase coherence exists even for
the long-wavelength fluctuations that control the transition.  Some
quantum critical points can be understood via mapping to standard
classical critical points in one higher dimension, but many of the
most experimentally relevant quantum critical points do not seem to
fall into this category.  Furthermore, even quantum critical points
that can be studied using the quantum-to-classical mapping have
important universal features such as frequency-temperature scaling
that do not appear at finite-temperature critical points. \cite{sachdevbook}

\subsection{Random systems and infinite randomness phases}

The properties of quantum critical points are often modified
dramatically by introducing randomness in their Hamiltonians.
Furthermore, the lower the dimensionality of the system, the more
dramatic the effect of disorder is. Several examples will be discussed in this review. Anderson
localization is the most Historically celebrated example; Anderson
has shown that a random chemical potential can completely localize an
otherwise propagating band \cite{Anderson} (see Sec. \ref {Anderson}). 

Another class of disorder induced phases was found when considering the random spin-1/2
Heisenberg chain \cite{MaDas1979,DotyFisher,BhattLee,DSF94,monthusreview}. A conclusive
solution of the ground state was found by D. S. Fisher, \cite{DSF94} who showed that any
amount of disorder will drive a spin-1/2 Heisenberg chain (as well as an
easy-plane xxz chain, and an xx chain) to the
random-singlet phase. As we discuss below (Sec. \ref{RSsec}), in this
phase, pairs of strongly interacting spins form singlets, which may
span arbitrarily large distances.
The random singlet phase is essentially a localized phase, with the
{\it typical} spin-spin correlations decaying as $C^{typ}(L)\sim
e^{-c\sqrt{L}}$. The typical correlations can be defined
as the exponent of $\overline{\ln \langle\hat{S}^z(0)\hat{S}^z(L)\rangle}$
(where an overline denotes disorder average). But if instead
we consider the {\it average} correlations, we find that the average
$\overline{\langle\hat{S}^z(0)\hat{S}^z(L)}\rangle$ is dominated by the
rare event of the two sites $0$ and $L$ being bound to a singlet. The
probability of this is essentially geometric, and falls off as
$1/L^2$. Therefore, $\overline{\langle\hat{S}^z(0)\hat{S}^z(L)\rangle}\sim
1/L^2$ as well. \cite{DSF94,DSF95,DSF98,RefaelFisher}  The random singlet phase
also arises as a localization problem of Majorana, as well as Dirac
Fermions, as demonstrated in Refs. \cite{YangBonesteel}.

The RS phase is but one example of an {\it infinite randomness fixed
point}. These points possess unique scaling
properties. Contrary to the standard
energy-length scaling at pure critical points, where $E\sim 1/L^z$,
the infinite randomness fixed points are also gapless, but with
the energy of an excitation scaling as:
\be 
\ln 1/E\sim L^{\psi}, 
\label{sc1}
\ee
Furthermore, the low energy
behavior of infinite-randomness fixed points is described by random
Hamiltonians with universal coupling distributions:
\be
\rho(J)\sim \frac{1}{J^{1-\chi/\G}},
\label{sc2}
\ee
$\G=\ln \Omega_0/\Omega$ is a parameter which keeps track of the energy
scale at which we probe the chain $\Omega$, relative to its largest bare
coupling $\Omega_0$.  $\chi$ and $\psi$ are universal constants, which
at the  random singlet
phase take the values $\psi=1/2$,  $\chi=1$. Other infinite randomness
fixed points maintain the form of the scaling, Eqs. (\ref{sc1}, \ref{sc2}),
but with $\psi=1/(n+1)$ and $\chi=n$ for other integer $n$ (these
numbers describe the Damle-Huse hierarchy; other combinations of
$\psi$ and $\chi$ may be possible).

The infinite-randomness fixed
points are, loosely speaking, the random analogs of pure CFT's, and
therefore it is important to understand all that we can about their
special universal properties, such as their entanglement
entropy. Generically, such points can be reached as instabilities
to disorder of well-known CFT's (e.g. in the XX, Heisenberg, and transverse-field Ising
model). Also, as we shall review, random gapless systems exhibit RG
flow between different infinite randomness fixed points. The study of
the entanglement entropy in these systems
allows us to explore any correspondence they may have with the pure
CFT c-theorem. Namely, does the entanglement entropy, or a related
measure, also decreases along flow lines of systems with randomness?
This question can be broken into two: (a) does the entanglement
entropy decrease along flows between pure CFT's and
infinite-randomness fixed points? (b) does the entanglement entropy
decrease along flows between two different infinite randomness fixed
points?

We note here that infinite randomness physics also has higher
dimensional analogs, as explored in Ref. \cite{BhattLee}, and
especially by several groups \cite{MotrunichHuse, LudwigFoster,
  mudry}. While we concentrate mostly on disordered systems in 1d,
Sec. \ref{higherD} will review relevant work in higher dimensions.

\section{Field guide to infinite-randomness fixed points}

One can not truly appreciate infinite-randomness criticality
without working through an example for each of the universality
classes mentioned above. As of now, all known infinite-randomness universality classes belong to the Damle-Huse series with $\psi=1/(n+1)$ and $\chi=n$\cite{DamleHuse}. The
simplest systems which give rise to the Damle-Huse series are
antiferromagnetic spin chains with $S=n/2$. Below we first review the
random-singlet phase ($n=1$) in the spin-1/2 Heisenberg model, and
then proceed to analyze how $n>1$ infinite-randomness critical points
arise in spin-$n/2$ chains, and in non-abelian anyonic models.

\subsection{Random-singlet phases - the simplest infinite randomness
  fixed points \label{RSsec}} 

The original infinite-randomness fixed point is the random
singlet phase. This phase describes the ground state of a spin-1/2 antiferromagnetic
Heisenberg chain with essentially any disorder distribution
\footnote{Excluding freak distributions such as $\rho(J)\sim (J
  \ln^{\alpha} J)^{-1}$ with $\alpha>1$.}.
Without disorder, the low-energy behavior of the spin-1/2 Heisenberg chain is described by a
conformal field theory with central charge $c=1$. Upon introduction of
disorder, the low-energy behavior of the chain flows to a different
critical phase: the random-singlet phase. \cite{BhattLee,DotyFisher,DSF94}

Let us write the random Heisenberg chain Hamiltonian:
\be
\H=\summ_{i=1}^{L} J_i \hat{\bf S}_i\cdot\hat{\bf S}_{i+1}
\ee
Roughly speaking, the strongest bond in the chain, say, $J_i$,
localizes a singlet between sites $i$ and $i+1$. Quantum fluctuations,
treated within second-order perturbation theory,
induce a new term in the Hamiltonian which couples sites $i-1$ and
$i+2$:\cite{MaDas1979,MaDas1980}
\be
\H'_{i-1,\,i+2}=\frac{J_{i-1}J_{i+1}}{2J_i} \hat{\bf S}_{i-1}\cdot\hat{\bf S}_{i+2}.
\label{MaDas}
\ee
Eq. (\ref{MaDas}) is the Ma-Dasgupta rule for the renormalization of
strong bonds. 

\begin{figure}\begin{center}
\includegraphics[width=7cm]{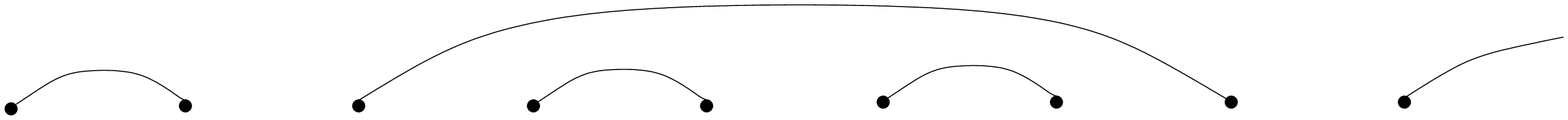}
\caption{The ground state of the random Heisenberg model consists of
  singlets connecting spins over arbitrarily long length
  scales.(figure taken from Ref. \cite{RKF} \label{RS_12}}
\end{center}
\end{figure}

This simple observation opens the way for an iterative
real-space renormalization group approach to random chains: we
identify the strongest bond in the chain, put the two spins it
connects into a singlet, and rewrite the Hamiltonian using the
Ma-Dasgupta rule (\ref{MaDas}) without the two recently singleted
spins, and with an effective and suppressed Heisenberg interaction between their
neighbors. After carrying out this iteration many times, the active
spins are a dilute irregular array compared to the initial
chain. Therefore the singlets they form occur over large length
scales, which increase as the decimation procedure progresses. The
solution is complete when all spins have been paired into singlets,
which automatically implies that the largest singlet connects spins
separated by a length comparable with the size of the system. A sketch of this phase is given in
Fig. \ref{RS_12}.

A useful parametrization of the couplings in the analysis of the random spin-1/2 Heisenberg chain is:
\be
\beta_i=\ln\frac{\Omega}{J_i}
\label{betadef}
\ee
where $\Omega$ is the highest energy in the Hamiltonian:
\be
\Omega={\rm max}_i \{J_i\},
\ee
and plays the role of a UV cutoff. The advantage of this
parametrization becomes clear when considering the Ma-Dasgupta rule in
these variables:
\be
J_{eff}=\frac{J_L J_R}{2\Omega}\rightarrow
\beta_{eff}=\beta_L+\beta_R-\ln 2
\label{MaDas1}
\ee
with $L$ and $R$ indicating bond to the left and to the right of the
decimated bond.  It is beneficial to define also a
logarithmic RG flow parameter:
\be
\G=\ln \frac{\Omega_0}{\Omega}
\label{gammadef}
\ee
where $\Omega_0$ is an energy scale of the order of the maximum
$J_i$ in the bare Hamiltonian. In terms of these variables, and using
the Ma-Dasgupta rule, Eq. (\ref{MaDas1}), we can construct a flow
equation for the distribution of couplings $\beta_i$:
\be
\frac{dP(\beta)}{d\G}=\der{P(\beta)}{\beta}
+P(0)\intt_0^{\infty}d\beta_1\intt_0^{\infty}d\beta_2
\delta_{\l(\beta-\beta_1-\beta_2\rr)}P(\beta_1)P(\beta_2).
\label{halfflow}
\ee
Here the first term describes the reduction of $\Omega$, and the
second term is the application the Ma-Dasgupta rule, where we neglect
$\ln 2$ in comparison to the $\beta$'s. For the sake of
readability, we denote the convolution with the cross sign:
\be
P(\beta_1)\stackrel{\beta}\times R(\beta_2)=\intt_0^{\infty}d\beta_1\intt_0^{\infty}d\beta_2
\delta_{\l(\beta-\beta_1-\beta_2\rr)}P(\beta_1)R(\beta_2).
\label{crossdef}
\ee

Eq. (\ref{halfflow}) has a simple solution, found by Fisher, which is an attractor to
essentially all initial conditions and distributions: \cite{DSF94}
\be
P(\beta)=\frac{\chi}{\G}e^{-\chi \beta/\G}.
\label{RSdist}
\ee
with $\chi=1$.

The function in Eq. (\ref{RSdist}) is the key to physical
characteristics of the random-singlet phase. For instance, to obtain the
energy-length scaling we make the following observations. Each
decimation step removes two sites, and the probability for a site to
be removed at RG scale $\Gamma\rightarrow\Gamma+d\Gamma$ is
$P(\beta=0)$. Therefore, the density of free spins evolves as:
\be
\frac{dn}{d\Gamma}=-2 n P(0)=-2 \frac{n}{\Gamma}
\ee
which is solved by:
\be
n=\frac{n_0}{\Gamma^2}
\ee 
upto an offset of order 1 in $\Gamma$. The length of a singlet formed
at RG scale $\Gamma$ must be of order of the average distance between
sites:
\be
\ell\sim 1/n.
\ee
This can be put into the length-energy scaling:
\be
\ell^{\psi}\sim \G=\ln 1/E
\label{LEscaling}
\ee
with $\psi$ being a universal critical exponent:
\be
\psi=1/2.
\ee

We note that almost exactly the same analysis applies to the
easy-plane xxz chain: \cite{DSF94} 
\be
\H=J\summ_i
\l(\hat{S}^x_i\cdot\hat{S^x}_j+ \hat{S}^y_i\cdot\hat{S^y}_j+\lambda
\hat{S}^z_i\cdot\hat{S^z}_j\rr) 
\label{xxz} 
\ee 
with $-1/2<\lambda< 1$. Under real-space RG, the anisotropy
parameter $\lambda$ flows to zero, which implies that the xxz
random-singlet phase is a fixed point distinct from its Heisenberg
model analog at the isotropic $\lambda=1$ point. 

\subsection{Transverse field Ising model}

\begin{figure}
\begin{center}
\includegraphics[width=7cm]{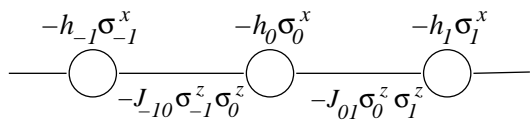}
\caption{The Hamiltonian of the transverse field Ising model. Each
site is  a spin-1/2 that interacts via Ising exchange  with its
nearest  neighbors and can be flipped by  the  local $x$-magnetic
field. (Figure taken from Ref. \cite{RefaelFisher}  \label{Hfig}} \end{center}\end{figure}

In addition to the Heisenberg model, early studies of infinite
randomness phases concentrated on the transverse field Ising model (TFIM)
\cite{DSF95}. 
The Hamiltonian of the random quantum Ising model is 
\be
\H=-\sum\limits_{i}\left(J_{i\,i+1}\sz{i}\sz{i+1}+h_{i}\sx{i}\right) 
\label{eq:tfisc} 
\ee 
with each site having two states, $\sz{}=\pm1$,  with quantum
fluctuations between them caused by the transverse, $\sx{}$,
fields. The system is illustrated in Fig. \ref{Hfig}.  The model has a
global symmetry of inversion about the $xy$ spin plane; breaking this
symmetry with $z$ fields would change the low energy
physics radically.

The analysis of the random TFIM follows the analysis of
the random Heisenberg model very closely, using real-space decimations of the
strongest terms in the Hamiltonian. The Hamiltonian consists of two
types of operators, $\sx{i}$ and $\sz{i}\sz{i+1}$, both of which have
eigenvalues $\pm 1$. Therefore it is natural to compare their
coefficients, $h_i$ and $J_i$ directly. The iterative decimation step
starts with finding the strongest coupling from the set
$\{h_i,J_i\}_{i=1}^{N}$. If $h_i=\Omega$ is the largest coupling, the
zeroth order solution of the wave function (i.e., ignoring all other
couplings in the Hamiltonian) will be an eigenstate of
$\sx{i}$ obeying $\langle\sx{i}\rangle=1$. Second order perturbation
analysis of the Ising bonds to the left
and right of site $i$ produces an Ising coupling:
\be
\H_{i-1,i+1}=\frac{J_{i-1}J_i}{\Omega}\sz{i-1}\sz{i+1}. 
\label{Jdec}
\ee
The coefficient of this Ising coupling clearly obeys
$J_{eff}<\Omega,\,J_i,\,J_{i-1}$.
 This is the TFIM analog of the
Ma-Dasgupta decimation step.
Similarly, if the strongest coupling is $J_i=\Omega$, the zeroth order
solution of the wave function will be an eigenstate of
$\sz{i}\sz{i+1}$ with $\langle\sz{i}\sz{i+1}\rangle =1$. There are two
states like that: $\sz{i}=\sz{i+1}=\pm 1$. The degeneracy
is lifted through quantum fluctuations induced by the terms
$-h_i\sx{i}-h_{i+1}\sx{i+1}$. Using second order perturbation theory we
obtain:
\be
\H_{i,i+1}=\frac{h_{i}h_{i+1}}{\Omega}\sx{i}\sx{i+1}=\frac{h_{i}h_{i+1}}{\Omega}\sx{i,i+1}. 
\label{hdec}
\ee
Eqs. (\ref{Jdec}, \ref{hdec}) are the TFIM analogs of the
Ma-Dasgupta decimation step for the Heisenberg model.

The decimation steps of the TFIM are also simplified by using
logarithmic variables for the Ising coupling and for the transverse
field. We define:
\be
\beta_i=\ln\frac{\Omega}{J_i}\hspace{10mm}\zeta_i=\ln\frac{\Omega}{h_i}\hspace{10mm}
\label{TFIM1}
\ee
and also follow the definition for $\G$, Eq. (\ref{gammadef}). The
decimation steps then become additive. For a field decimation at site
$i$, we have:
\be
\beta_{eff}=\beta_{i}+\beta_{i+1}.
\ee 
For a bond decimation, the
effective field becomes 
\be
\zeta_{eff}=\zeta_{i}+\zeta_{i+1}.
\ee
 
The TFIM at low energies is characterized by coupling and transverse
field distributions $P(\beta)$ and $R(\zeta)$ respectively, which obey
the following flow equations:
\be
\ba{c}
\frac{dP(\beta)}{d\G}=\der{P(\beta)}{\beta}
+R(0)P(\beta_1) \stackrel{\beta}\times   P(\beta_2)+P(\beta)(P(0)-R(0)),\vspace{2mm}\\
\frac{dR(\zeta)}{d\G}=\der{R(\zeta)}{\beta}
+P(0)R(\zeta_1)\stackrel{\zeta}\times R(\zeta_2)+R(\zeta)(R(0)-P(0)).
\label{TFIMflow}
\ea
\ee

The quantum Ising model exhibits a quantum phase transition in its ground state when the 
nearest neighbor interaction and the transverse field are of comparable
strength. In a non-random model this occurs when $J=h$ \cite{sachdevbook}.  
In a random system, where the $J$'s and the $h$'s are drawn independently from some  distributions,  a solution of the flow equations
(\ref{TFIMflow}) shows that the transition occurs when 
$\overline{\log h}=\overline{\log J}$, where the over-bars denote
averaging over the randomness \cite{DSF95}. 
A convenient parametrization of the proximity to the transition is   
\be
\delta\equiv\frac{\overline{\log h_I}-\overline{\log J_I}}{var(\log h)+var(\log J)}
\label{eq2}
\ee
with $\delta>0$ yielding the disordered phase, and $\delta<0$ yielding
the ferromagnetic phase, where a large cluster forms due to the Ising interaction. Using this
parametrization, Fisher found that at low energies, at the fixed
point, $\delta=0$, and at the Griffiths phase
that surrounds it, the distribution functions are given by:
\be
\hspace{-1cm}
P(\beta)=\frac{\delta e^{-\delta\Gamma}}{2\sinh\delta\Gamma}e^{-(-\delta+\delta\coth(\Gamma\delta))\beta}
\hspace{0.5cm}
R(\zeta)=\frac{\delta e^{\delta\Gamma}}{2\sinh\delta\Gamma}e^{-(\delta+\delta\coth(\Gamma\delta))\zeta}.
\label{TFIMsol}
\ee

As can be easily seen, when $\delta=0$, the two distributions for
coupling and transverse field become identical, and the low energy
behavior of the random TFIM obeys the same infinite randomness scaling
as the random singlet phase. The picture of the low energy phase,
however, is quite different. During the renormalization process, as
the energy scale is reduced, cluster of parallel spins form and grow
to length scale $\ell$, and then they freeze in a superposition of
pointing up and down as clusters, i.e., in the x-direction of the
collective spin. The excitation energy of of such cluster scales as 
$e^{-\sqrt{\ell}}$. Entanglement entropy resides in the cluster
formation: the entanglement entropy between two sections is equal to
the number of clusters that connect them.

When $\delta$ is non-zero but small, the random TFIM is in a
Griffiths phase, where the low energy behavior
dominated by gapless but well localized excitations. Thinking about
the low energy behavior using the decimation picture, at early stages
the chain obeys the critical $\delta=0$ scaling. But when the typical cluster sizes and bond
lengths  are of  the same order of the {\it correlation length}
\be
\xi \approx \frac{1}{\delta^2},
\ee
and the log-energy scale is of order 
\be
\Gamma_\times\sim\frac{1}{\delta},
\ee
a crossover to the ordered ($\delta<0$) or disordered ($\delta>0$)
occurs. The energy-length scaling is then different from that at the critical point. For small $\delta$,
in both phases 
\be
\Omega\sim \ell^{-z(\delta)}
\ee
with the effective dynamical exponent,
\be
z\approx \frac{1}{|\delta|}
\ee
near the critical point.

\subsection{Infinite-randomness fixed points for higher spin}

As mentioned in the beginning of this section, the random-singlet phase, with universal distribution (\ref{RSdist})
and length-energy scaling, Eq. (\ref{LEscaling}) is but one example
of an {\it infinite randomness fixed point}. In general, disordered
systems may have a similar type of logarithmic length-energy scaling
relations, Eq. (\ref{LEscaling}), with different $\psi$ and $\chi$ in
Eq. (\ref{sc1}, \ref{sc2}). As mentioned above, to date, all known universality classes of infinite-randomness fixed points
can be realized in Heisenberg models with spin $s\ge 1/2$. Following
Refs. \cite{MGJ,HymanYang} and \cite{RKF} which dealt with the spin-1 and
spin-3/2 cases respectively, Damle and Huse showed that a spin-$s$
chain may exhibit infinite-randomness fixed points with \cite{DamleHuse}:
\be
\ba{cc}
1/\psi=2s+1,\,& \chi=2s.
\ea
\ee
These fixed points were dubbed domain-wall symmetric fixed points.

\begin{figure*}
\begin{center}
\includegraphics[width=10cm]{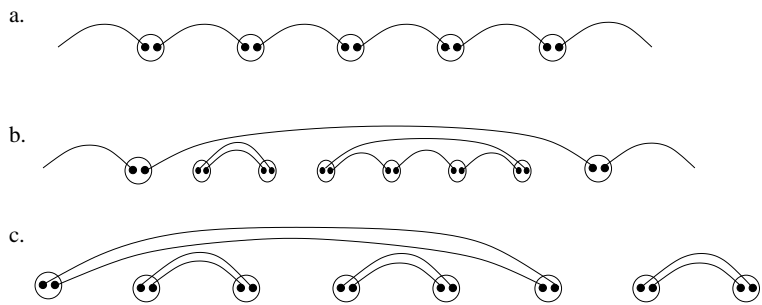}
\caption{(a) At very low disorder, the ground state of the spin-1
  Heisenberg chain is well described as a valence-bond solid. Each
  spin-1 site is described by two spin-1/2 parts (black dots) that are
  symmetrized. Each site forms a spin-1/2 singlet to its right and to
  its left. (b) As disorder grows, defects appear in the VBS
  structure, and the gap is suppressed. (c) At very high disorder, a
  phase transition occurs to the spin-1 random singlet (RS) phase. 
 \label{VBS}}
\end{center}
\end{figure*}

\subsubsection{Domain-wall picture}

The simplest illustration of the domain-walls picture is in the
spin-1/2 random singlet phase. The random-singlet
phase forms through a competition between two possible singlet
domains: Domain (1,0) with singlets appearing on odd bonds only, and
domain (0,1) with singlets appearing on even bonds. Generalizing this
concept to higher spins is straightforward: a spin-$s$ site can be
represented as $2s$ spin-1/2 parts with a permutation symmetric
wavefunction. The notation $(a,2s-a)$ with $0\le a\le 2s$ then signifies a domain with $a$ spin-1/2
singlet links on odd bonds, and $2s-a$ spin-1/2 singlet links on even
bonds. Each singlet link puts {\it one} of the spin-1/2 parts in
neighboring sites in a singlet. Having $a$ such singlet-links between
two spin-$s$ sites constrains their total spin to be $s_{total}\le
2s-a$ \cite{AKLT}. 

\begin{figure*}\begin{center}
\includegraphics[width=12cm]{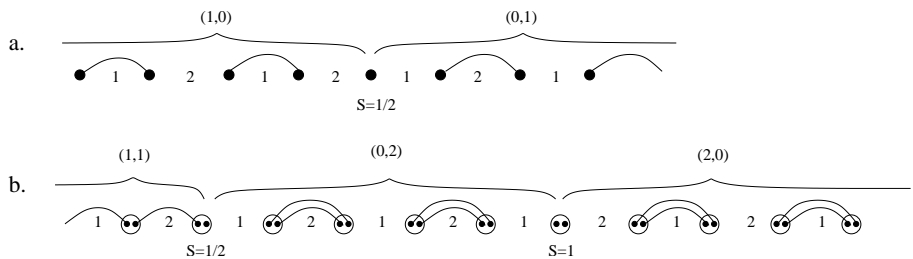}
\caption{(a) Two domains are possible
 in the spin-1/2 Heisenberg model - (1,0) and (0,1). A domain wall between them gives rise to a spin
  1/2 effective site. (b) In the case of a spin-1 chain, there are
  three possible domains: (1,1), (2,0), and (0,2), domain walls
  between them are effectively a spin-1/2 and spin-1 sites respectively. 
\label{domainwall}}
\end{center}\end{figure*}

This notation makes it easy to think about randomness as
competition between different dimerizations. For each
domain, there is a probability $\rho_a$ to be of type $(a,2s-a)$, and
also, for each domain, there is a transfer matrix, which tells the
probability of domain $a$ to be followed by domain $a'$, which is
$W_{aa'}$. Note that:
\[
\summ_{a'=0}^{2s}W_{aa'}=1.
\]
At any finite temperature or energy scale, the
non-frozen degrees of freedom (i.e., spins that were not yet
decimated) lie on domain walls. Thus in the domain wall between the (1,0)
and (0,1) domains, there is one free spin-1/2 site (see
Fig. \ref{domainwall}a). This free spin interacts with similar
spin-1/2's in neighboring domain walls through an interaction mediated
by quantum fluctuations of the domain in between. Thus each domain of
type $a$ is
associated with a bond between neighboring free spins, and has a
distribution of coupling $P_a(\beta)$, with $\beta$ defined in
Eq. (\ref{betadef}). 

The renormalization of strong bonds is now
described as the decimation of domains. In the spin-1/2 chain,
whenever a domain is decimated, its two neighboring domains, being identical, unite to
form a single large domain; thus a singlet appears over the decimated domain, and
connect the spins on the two domain walls. This is the
Ma-Dasgupta decimation step, Eq. (\ref{MaDas}). The random-singlet phase appears when the
(1,0) and (0,1) domains have the same frequency. It is a
critical point between the two possible dimerized phases associated
with the two domains.

In $s>1/2$ spin chains, the domain picture is richer. In the spin-1
Heisenberg chain there are  three possible domains: (0,2), (1,1), and
(2,0). The VBS is associated with the (1,1) phase, which has a uniform
covering of the chain with spin-1/2 singlet links. On the other hand, the strong
randomness random-singlet phase in this system occurs when the
competing domains are (2,0) and (0,2). This is completely analogous to
the spin-1/2 random singlet phase, except that the domain walls
consist of free spin-1 sites (see Fig. \ref{domainwall}b). 

A general
domain wall between domains $a$ and $a'$ can be easily shown to have
an effective spin:
\be
S_e=\frac{1}{2}|a-a'|
\ee
as each singlet link leaving the domain wall removes a spin-1/2 from
it.

Typically, the decimation of a domain involves forming as many singlet
links as possible between the two domain walls. If the two neighboring
domains are identical, $a'=a''$, then so are the domain walls,
and a full singlet is formed; this is the Ma-Dasgupta decimation rule in Eq. (\ref{MaDas}). If
the two domain walls are not identical, and interact with each other
anti-ferromagnetically, singlet links forming between the two domain
walls will exhaust one of the domain-wall spins, and the domain
$D_{a}$ will be swallowed by the domain containing the exhausted spin.

If the two domains neighboring a strong bond are different, $a'\neq
a''$, and the interaction between the two domain-wall spins is
ferromagnetic, the two spins unite into the domain wall between
$D_{a'}$ and $D_{a''}$. For example, consider a (1,1) domain with an even number of links. it
has to connect between a (2,0) domain and a (0,2) domain; both domain
walls will have spin-1/2. Upon
decimation of the (1,1) domain, we are left with a domain wall between
(2,0) and (0,2), which has a spin-1.  

Indeed at the critical point between the Haldane and
random-singlet phases all three domains appear with equal probability--hence
the designation ``permutation-symmetric critical point''. Since each domain appears with the same
frequency at the critical point, each possible domain wall appears with the same
frequency as well. Two domain walls: $(0,2)-(1,1)$ and $(2,0)-(1,1)$ are 
effective spin-1/2's, whereas the third possible domain wall,
$(2,0)-(0,2)$ is a spin-1. Thus, at any finite but low temperature or
energy scale, 2/3 of the unfrozen degrees of freedom are
effectively spin-1/2, and 1/3 are spin-1. These fractions are
universal and a direct consequence of the bare spin of the model. 

In general, each domain will have a logarithmic coupling strength
$\beta$, and a distribution function $P_a(\beta)$. The flow equations
for the transfer matrix probability $W_{aa'}$
are \cite{DamleHuse}:
\be
\ba{c}
\frac{dW_{aa'}}{d\G}=V_{aa'}-\frac{1}{2}W_{aa'}\l(P_a(0)+P_a'(0)-V_{aa}-V_{a'a'}\rr)\vspace{2mm}\\
V_{aa'}=\summ_{b} W_{ab}P_b(\beta)W_{ba'}\\
\ea
\label{DWflow0}
\ee
The flow equations of the distribution functions are:
\be
\frac{dP_a(\beta)}{d\G}=\der{dP_a(\beta)}{d\beta}+P_a(0)P_a(\beta)+V_{aa}\l(P_a(\beta_1)\stackrel{\beta}\times
P_a(\beta_2)-P_a(0)\rr).
\label{DWflow}
\ee
The perturbation symmetric fixed point solution of the above equations
has all domains being equivalent: they appear with the same probability
and have the same distribution function (note that this is only the
fixed-point solution, and on the critical manifold the bare system
need not exhibit this symmetry; it is emergent). Eqs. (\ref{DWflow0}, \ref{DWflow}) are then
solved by the attractive solution:
\be
P_a(\beta)=\frac{2s}{\G} e^{-2s\beta/\G},
\label{s11}
\ee
\be
\rho_a=\frac{1}{2s+1}
\label{s12}
\ee
and:
\be
W_{aa'}=\frac{1}{2s}(1-\delta_{aa'})=\frac{1}{2}(1-\delta_{aa'}).
\label{s13}
\ee
Eqs. (\ref{s11}-\ref{s13}) give a complete description of the spin-s
critical point, and lead to the energy-length
scaling properties, Eq. (\ref{LEscaling}), of:
\be
L\sim \frac{1}{\G^{2s+1}},
\label{LG}
\ee
i.e.,
\be
L^{1/(2s+1)}\sim \ln \Omega_0/\Omega.
\ee
For the spin-1 case, in particular, the permutation symmetric critical
point describes the critical point between the Valence-Bond-Solid (VBS) Haldane
phase, and the Random Singlet phase of a spin-1 Heisenberg chain.

\subsection{Infinite-randomness fixed points of non-abelian anyons \label{non-abelian}}.

Motivated by quantum Hall physics \cite{ReadRezayi,ReadRezayi2,SternReview}, and their possible
application in topological qubits \cite{Kitaev,Kitaev2, NayakReview}, the study of interacting
non-abelian anyonic systems has moved to center stage. As
non-abelian anyons are expected to appear as defects in quantum Hall
states such as $5/2$ or $12/5$ \cite{ReadRezayi,ReadRezayi2}, it is natural to ask how
a disordered system of such anyons behaves. The study of random
non-abelian chains as started with the consideration of the
random-singlet phase of Majorana fermions, and Fibonacci anyons
\cite{YangBonesteel,FRBM}, and continued with the study of infinite
randomness fixed points in the more general class of non-abelian
chains, the truncated $SU(2)_k$ systems\cite{FLTR}. 

Non-abelian anyons are characterized first and foremost by fusion
rules. A trivial example of a fusion rule in a simple abelian system
is given by considering two spin-1/2's sites, which can fuse
according to the the SU(2) rule:
\be
\frac{1}{2}\otimes\frac{1}{2}=0\oplus 1.
\label{FRhalf}
\ee
The random singlet phase of the AFM Heisenberg model arises when we
always choose to fuse strongly interacting neighbors into the singlet
(spin-0) state. For non-abelian anyons, the spin-compounding rule, Eq. (\ref{FRhalf}) is substituted by the {\it
  fusion algebra} of the non-abelian system:
\be
a\otimes b =\oplus\summ_c N^{c}_{ab} c,
\label{FA}
\ee
where $N_{ab}^{c}$ is the number of ways the superselection sectors
$a$ and $b$ can fuse into $c$. 

A major difference, however, between rules (\ref{FRhalf}) and
(\ref{FA}) is that fusion rules for a non-abelian algebra are always {\it
  closed}, while in regular spin-chains, the fusion rules include an
infinite set of subspaces. The closure of the fusion rules results
from the nonlocality of the Hilbert space of non-abelian systems. 
 It implies that
one can {\it always} apply a real-space RG scheme without ever
generating new types of coupling in the renormalized
Hamiltonian.  Furthermore, just as in conventional spin chains, a
decimation will result either in a Ma-Dasgupta renormalization of the
neighboring couplings, or in their multiplication by a factor of
magnitude smaller than 1. Therefore {\it sufficiently disordered (and most
likely even weakly disordered) non-abelian chains will exhibit an
infinite randomness behavior in the large length scale properties of
their ground state.}  Another counter-intuitive consequence of the fusion algebra, Eq. (\ref{FA}) is that
the Hilbert space of individual non-abelian anyons (which are part of
an interacting non-abelian system) can have non-integer
dimension. 
 
An important class of non-abelian anyons is the $SU(2)_k$
algebra, which arises in Read-Rezayi quantum Hall states that may describe fillings
$\nu=n\pm k/(2+k)$ \cite{ReadRezayi}. This algebra is the truncated $SU(2)$ algebra,
which allows two spins (corresponding to quasiparticles of the
Read-Rezayi states)
 $s_1,\,s_2\le k/2$ to fuse into objects of
total spin $s_{total}\le k/2$. The fusion rule is then:
\be
s_1\otimes s_2=|s_1-s_2|\oplus \ldots \oplus
\textrm{min}\{s_1+s_2,k-s_1-s_2\}
\ee
When $s_1=s_2$, the two spins can fuse into the singlet state, i.e.,
the identity, ${\bf 1}$. 

Let us now consider a chain of $N$ $SU(2)_k$ spin-1/2's. The dimension of the
Hilbert space is given by ${\rm dim}\, H_N\approx d^N$  with
\be
d=2\cos\l(\frac{\pi}{2+k}\rr).
\ee
The random Heisenberg AFM then has the form
\be
\H=\summ_i J_i P_{i,\,i+1}^{\bf 1},
\label{NAH} 
\ee
where $ P_{i,\,i+1}^{\bf 1}$ is the projection operator of two
sites onto the identity subspace. When all couplings $J_i$ are the
same, the low energy behavior of the $SU(2)_k$ chain is described by a
CFT with central charge $1-6/[(k+1)(k+2)]$. Clearly, when the $J_i$'s
are random, this Hamiltonian is
amenable to the Ma-Dasgupta decimation procedure. By applying second
order perturbation theory using the F-matrix formalism of tensor
categories, Bonesteel and Yang find that when two sites are bound into
a singlet state, their neighbors interact with strength
\cite{YangBonesteel}:
\be
J_{eff}=\frac{2}{d^2}\frac{J_L J_R}{J_m}
\label{NAMD}
\ee
with $J_{L/R}$ are the couplings to the left and right of the
decimated bond, whose strength is $J_m$. Since $2/d^2<1$, the
$SU(2)_k$ Heisenberg model always flows to the random singlet fixed
point, with arbitrarily weak randomness.

\subsubsection{Majorana fermions.}

The simplest example of a random-singlet phase in a non abelian system
occurs in a  Majorana Fermion chain \cite{YangBonesteel}. Majorana fermions, or real
fermions, as they are sometimes called, can be constructed from
fermion creation and annihilation operators,
$\hat{\psi}^{\dagger},\hat{\psi}$ \cite{Wenbook}. We can construct two anti-commuting
Majorana operators:
\be
\sigma_1=\hat{\psi}^{\dagger}+\hat{\psi}\hspace{0.5cm}
\sigma_2=i(\hat{\psi}^{\dagger}-\hat{\psi})
\label{mtrans}
\ee
note that $\{\sigma_i,\,\sigma_j\}=2\delta_{ij}$, and that
  $\sigma_i=\sigma_i^{\dagger}$. Majorana fermions arise in the $SU(2)_2$
algebra, which also describes the CFT of the 2d Ising model. The fact
that two Majorana fermions are required to form a fermionic state is
reflected in the fusion algebra of $SU(2)_2$:
\be
\sigma\otimes \sigma ={\bf 1}\oplus\psi,\\
\psi \otimes \sigma= \sigma,\\
\psi\otimes\psi={\bf 1}.
\label{tauFA}
\ee
$\psi$ is a chiral fermionic state which
arises when two $\sigma$'s are combined; it can either be occupied or
empty. This double degeneracy of pairs of Majorana discloses their
quantum dimension:
\be
d=2\cos(\pi/4)=\sqrt{2}.
\ee

An $SU(2)_2$ Heisenberg chain consists of an array of $\sigma$, which
are the spin-1/2, quasiparticles:
\be
\H=\summ_j J_j i \sigma_j \sigma_{j+1}.
\label{MH}
\ee
For the Ma-Dasgupta procedure, we need to find the eigenstates of
a single bond, which is readily done by using Eq. (\ref{mtrans}):
\be
\H=J_{2n}i\sigma_{2n}\sigma_{2n+1}=J_{2n}(2\hat{\psi}_{n}^{\dagger}\hat{\psi}_{n}-1).
\ee
Thus, the ground state of the $2n$ bond corresponds to an empty
fermionic state, and the excited state is the filled state. Using
second order perturbation theory in conjunction with
Eq. (\ref{mtrans}) yields Eq. (\ref{NAMD}), and the random singlet
ground state of the Majorana chain. 

It is instructive to note that by associating $\sigma_{2n}$ and
$\sigma_{2n+1}$ with the fermionic state $\hat{\psi}_n$, we can
rewrite Eq. (\ref{MH}) as:
\be
\hspace{-2cm}\H=\summ_n \l[J_{2n}\l(\hat{\psi}_{n}^{\dagger}\hat{\psi}_{n}-\hat{\psi}_{n}\hat{\psi}_{n}^{\dagger}\rr)+J_{2n+1}\l(\hat{\psi}_{n}\hat{\psi}_{n+1}^{\dagger}-\hat{\psi}_{n}^{\dagger}\hat{\psi}_{n+1}+\hat{\psi}_{n}\hat{\psi}_{n+1}-\hat{\psi}_{n}^{\dagger}\hat{\psi}_{n+1}^{\dagger}\rr)\rr].
\label{MH-tfim}
\ee
which is the Hamiltonian for the TFIM after a Wigner-Jordan
transformation. Intuitively, the Jordan-Wigner transformation maps the
Ising bonds into the fermionic states, and the transverse field into
the odd bonds Majorana interaction, which results in hopping and
pairing terms for the Fermionic states.

\subsubsection{Fibonacci anyons.} 

Fibonacci anyons arise in the $SU(2)_3$ algebra, which may describe
the $\nu=12/5$ quantum Hall state \cite{ReadRezayi, ReadRezayi2}. These non-abelian
anyons derive their name from the fusion algebra
\be
\tau\otimes \tau ={\bf 1}\oplus\tau.
\label{tauFA1}
\ee
If we consider a chain of $N$ Fibonacci anyons, and ask how big is the Hilbert
space they span, we can write for the combination of $n+1$ particles:
$d_{n+1}=d_{n-1}+d_n$, since whenever we combine another anyon it
either forms a singlet with the topological charge, or another $\tau$
anyon. This recursion is solved by:
\be
d_N\approx d^N=(\frac{1+\sqrt{5}}{2})^N,
\ee
i.e., the dimension $d$ of Fibonacci anyons is the golden ratio. 

\begin{figure}
\begin{center}
\includegraphics[width=7cm]{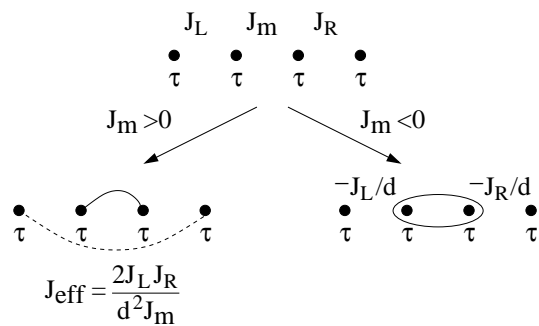}
\caption{The two real-space RG steps in the Fibonacci chain. If
  $J_m>0$, the Ma-Dasgupta AFM decimation of two $\tau$'s leads to an AFM bond
  between the nearest neighbor, as Eq. (\ref{NAMD}) prescribes. In the
  FM case, where $J_m<0$, two $\tau$'s fuse and {\it reverse} the sign
  of the interaction of the fused $\tau$ with its neighbors [see
  Eq. (\ref{FMdec})]. \label{FibRGfig}}
\end{center}
\end{figure}

The Fibonacci random Heisenberg model is identical to Eq. (\ref{NAH}),
but we can consider this model beyond the AFM model, by allowing $J_i$
to be both positive and negative. This results in the FM/AFM Fibonacci
model. A FM bond corresponds to a bond which is satisfied if the two
anyons it connects fuse into another anyon, i.e., to the non-trivial
fusion channel in rules (\ref{tauFA1}):
\be
J_i P_{i,\,i+1}^{\bf 1}=-J ( P_{i,\,i+1}^{\bf \tau}-1).
\ee
When addressing a random FM/AFM Fibonacci chain, we can still use the
real-space decimation procedure, with the Ma-Dasgupta rule,
Eq. (\ref{NAMD}) for AFM bonds, and with strong FM bonds renormalizing
two sites into a single Fibonacci site: $\tau_i\otimes
\tau_{i+1}=\tau_{i,\,i+1}$. By using the F-matrix rules for the
Fibonacci anyons, we arrive at the surprising renormalization rules of
neighboring bonds (see Ref. \cite{FRBM}): 
\be
J_{i\pm 1}\rightarrow -\frac{1}{d} J_{i\pm 1},
\label{FMdec}
\ee
where $d$ is the golden ratio (see Fig. \ref{FibRGfig}).

In order to find the fixed points of the Fibonacci anyons, we need to
write flow equations for both FM and AFM bond distributions, which we
denote $N(\beta)$ and $P(\beta)$ respectively. We obtain
\cite{FRBM}:
\be
\ba{c}
\frac{dP}{d\Gamma}=\der{P}{\beta}+P(0)\l(P \stackrel{\beta}\times P+N\otimes
N)\rr)+2N(0)N(\beta)-N(0)P(\beta)\\
\frac{dN}{d\Gamma}=\der{N}{\beta}+2P(0)N\stackrel{\beta}\times P-N(0)N(\beta)+2N(0)P(\beta),
\ea
\label{flow}
\ee
where we again use the notation: $F\stackrel{x}\times G=\intt_{0}^{\infty} dx_1 \intt_{0}^{\infty} dx_2
\delta(x-x_1-x_2) F(x_1)G(x_2)$. An exponential ansatz,
$N(\beta)=n_0 e^{n_0\beta/\G},\,P(\beta)=p_0 e^{p_0\beta/\G}$, for the
fixed point distributions reveals two solutions. First, the pure AFM fixed
point, which is just the random-singlet phase:
\be
\ba{cc}
n_0=0 & p_0=1.
\ea
\ee
But a second fixed point is found
by
\be
p_0=n_0=1.
\ee
This fixed point has an equal proportion of FM and AFM bonds, and
although it is an infinite-randomness fixed point, with coupling distributions:
\be
N(\beta)=P(\beta)=\frac{1}{\Gamma}e^{-2\beta/\Gamma}.
\ee
This new fixed point belongs to the spin-1 Damle-Huse universality
class, which describes the three-domain permutation symmetric fixed
point between the Haldane phase and the random singlet phase in a
random spin-1 Heisenberg chain \cite{DamleHuse,
  HymanYang,MGJ}. 

Surprisingly, a stability analysis reveals that the
random singlet fixed point is actually unstable to flow to the FM/AFM
mixed fixed point. This is a consequence of the FM decimation step,
Eq. (\ref{FMdec}), which allows a single FM bond to shift two neighboring AFM bonds to FM upon
decimation. A reasonable conjecture is that the pure FM Fibonacci
chain flows to the infinite-randomness mixed FM/AFM fixed point upon
disordering of the couplings $J_i$'s. The pure FM fixed point is
described by a CFT with central charge $c=4/5$ \cite{Qpeople1,Qpeople2}.

Another note is that
the mixed FM/AFM random Heisenberg chain for $SU(2)$ spin-1/2's was
studied in Refs. \cite{LeeSigrist1,LeeSigrist2}). Within the
real-space RG approach it is easy to see that higher and higher spins
are generated, and as a result, the model has a {\it finite}
randomness fixed point which was observed numerically.

\subsubsection{General $SU(2)_k$ models.} 

From the Fibonacci example it is easy to see how we can obtain
different Heisenberg models which are not purely AFM  for any
$SU(2)_k$ models. For $k$ odd, it was shown in Ref. \cite{FLTR}
that all fixed points arising in these models (i.e., nearest neighbor
random chains) are indeed infinite randomness fixed points. A bit
disappointingly, all of these fixed points belong to the Damle-Huse
hierarchy; the permutation symmetric points of spin-s Heisenberg
models are realized in $k=2s+1$ truncated $SU(2)_k$ models; some
intuition to this relationship is that the types of non-abelian anyons
are mapped into domains in the spin-models. The
topological properties of the $SU(2)_k$ algebra, however, guarantee
that the Damle-Huse points are stable fixed points, and therefore
describe {\it phases} rather than critical points. This indeed fits
our finding of the mixed FM/AFM fixed point of the Fibonacci chain ($k=3$)
being stable, and corresponding to the $s=1$ Damle-Huse fixed point.

\subsection{Generalized transverse field models}

A model that played an important role in the study and understanding of the entanglement entropy of
random spin chains is the generalized random transverse field model (GTFIM)
introduced by Santachiara \cite{Santachiara}. Each site in the $N$-flavor GTFIM has
$N$ states, $\{\ket{q}\}_{q=1}^N$. The Hamiltonian for this
model is:
\be
\H_N=-\summ_i J_i \summ_{n=1}^{N-1} \alpha_n \l({S^z_i}^{\dagger} S^z_{i+1}\rr)^n-\summ_i h_i \summ_{n=1}^{N-1}
\alpha_n \Gamma_i^n
\label{GTFIM}
\ee
with:
\be
S^z_i\ket{q}_i=e^{2\pi i q/N}\ket{q}_i,\, \Gamma_i =\summ_{q=1}^N
(\ket{q}_i {}_i\bra{q+1}+\ket{q+1}_{i} {}_i\bra{q})
\ee
with $\ket{N+1}=\ket{1}$. The coefficients $\alpha_n\ge 0$ obey
$\alpha_n=\alpha_{N-n}$ to maintain hermiticity of the model. 

To carry out a real space RG analysis we need to find the eigenstates
of the local operators in the Hamiltonian (\ref{GTFIM}). The eigenstates of the
$S^z_iS^z_{i+1}$ term in this Hamiltonian are of the type:
\be
\ket{q,\Delta q}_{i,\,i+1}=\ket{q}_i\ket{q+\Delta q}_{i+1}
\ee
with energies:
\be
E^{zz}_{\Delta q}=-\summ_{n=1}^{N-1}\alpha_n \cos\l(2\pi n \Delta q /N\rr)
\ee
with the ground states being N times degenerate, $\ket{q,\Delta
  q=0}_{i,\,i+1}=\ket{q}_i\ket{q}_{i+1}$. 
The eigenstates of the
$\Gamma_i$ term are:
\be
\ket{p}_{i}=\summ_{q=1}^{N}e^{2\pi i p/N}\ket{q}_i
\ee
with energies:
\be
E^{\Gamma}_{p}=-\summ_{n=1}^{N-1}\alpha_n \cos\l(2\pi n p/N\rr).
\ee
The fact that the spectrum of the $S^z S^z$ and $\Gamma$ terms is
identical is a reflection of a duality which the GTFIM possesses.

Let us go to the decimation procedure for the case of a strong $J$ and
$h$. When encountering a strong $J_i=\Omega$, we set $q_i=q_j=q$ to
form the $\ket{q,\Delta q=0}_{i,\,i+1}$ eigenstate. This eigenstate is
$N$ times degenerate. The degeneracy between the various $q$ states is
lifted through the action of the $\Gamma_i$ and $\Gamma_{i+1}$ terms,
which in second order perturbation theory form an effective term:
\be
\H_{i,i+1}=\frac{h_i h_{i+1}}{\kappa_1\Omega}
\summ_{n=1}^{N-1}\tilde{\alpha}_n \Gamma_{i,i+1}^n
\ee
with:
\be
\ba{c}
\kappa_n=\frac{1}{2}\summ_{m=1}^{N-1} \alpha_m \l( \cos(2\pi m
n/N)-1\rr)\\
\tilde{\alpha}_n=\alpha_n^2 \frac{\kappa_1}{\kappa_n}.
\ea
\label{kaRG}
\ee
The decimation step for a strong $h_i=\Omega$ entails setting site $i$
into the state $\ket{p=0}_i=\summ_{q=1}^{N}\ket{q}_i$. Through a
second order process, the bonds to
the left and right of the decimated site induce an effective
interaction:
\be
\H_{i-1,i+1}=\frac{J_{i-1} J_{i}}{\kappa_1\Omega}
\summ_{n=1}^{N-1}\tilde{\alpha}_n \l({S^z_{i-1}}^{\dagger}S^{z}_{i+1}\rr)^n
\ee
with $\kappa_n$ and $\tilde{\alpha}_n$ following Eq. (\ref{kaRG}).

The discussion above was completely general, and did not depend
on the actual values of $\alpha_n$. Therefore, under the condition
that $\alpha_n\le 1$, regardless of the choice of $\{\alpha_n\}$, a
GTFIM also flows to an infinite randomness fixed point, with the same
universal properties as the TFIM, but with a site degeneracy of $N$
rather than $2$.

An important realization of the GTFIM has the following choice of
$\{\alpha_n\}$:
\be
\alpha_n=\frac{\sin(\pi/N)}{\sin(\pi n/N)}.
\ee
With this choice of $\alpha_n$, the critical ($J=h$) pure GTFIM
coincides with the parafermionic $Z_n$ CFT, which has central charge:
\be
c_N=2\frac{N-1}{N+2}.
\label{para-c}
\ee
Upon allowing $J_i$ and $h_i$ vary randomly at each site, we can apply
RSRG to this model as well. After a few application of the
renormalization rules, and Eq. (\ref{kaRG}) in particular, the set
$\{\alpha_n\}$ renormalizes to:
\be
\tilde{\alpha}_n=\delta_{1,n}, 
\ee
which coincides with the $Z_N$ clock model. This model was analyzed in
Ref. \cite{SenthilMajumdar} and was shown to flow to strong
randomness as anticipated above.

\section{Entanglement entropy in infinite-randomness fixed points}

\subsection{Spin-1/2 random singlet entanglement entropy \label{RSent} }

\begin{figure}
\begin{center}
\includegraphics[width=8cm]{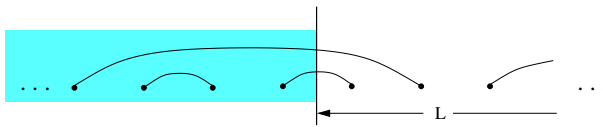}
\caption{The entanglement entropy of a segment is the
  number of singlets that connect the segment with the rest of the
  chain (shaded area). In this example there are two such singlets.
\label{RSfig2}} 
\end{center}
\end{figure}

The simplest case for computing the entanglement entropy is the
random singlet phase of the spin-1/2 Heisenberg model.
The entanglement entropy of a
spin-$1/2$ particle in a singlet with another such particle is $1$ (bit),
which is the entropy of a spin (e.g. $a$) in a singlet,
$\ket{\psi}=\frac{1}{\sqrt{2}} (\ket{\uparrow}_a \ket{\downarrow}_b-
\ket{\downarrow}_a \ket{\uparrow}_b)$ with its partner, $b$, traced out. The entanglement of a segment of the random
Heisenberg chain is just {\it the number of singlets that
  connect sites inside to sites outside the segment} (Fig. \ref{RSfig2}). 

To obtain the entanglement, we calculate the number, $N$, of singlets
that form over a single bond $B$ up to the length-scale $L$. A quick argument shows that the
entanglement should scale logarithmically with the length $L$. If we neglect the history
dependence of the distribution of bond $B$, we can find $N$ by using
the distribution of bond strengths, Eq. (\ref{RSdist}). The probability
of a singlet forming across the bond $B$ when we
change the energy scale $\Omega\rightarrow \Omega-d\Omega$,
$\G\rightarrow \G+d\G$ is also the average number of singlets
forming. It is given by:
\be
d\overline{N}=d\G P(\beta=0)=\frac{d\G}{\G}
\label{eq3.1}
\ee
integrating this leads to $\overline{N}=\ln \G\approx \frac{1}{2}\ln
L$ if we stop counting at length scale $L\sim \G^2$.

Taking into account that the entanglement of a segment of length $L$
with the rest of the chain has two ends, we get: 
\be
S_L\sim N_L\approx 2\cdot\ln \sqrt{L}+k=\ln L+k,
\label{eq3.3}
\ee
where $k$ is a non-universal constant, which also depends on the
initial realization of the disorder.


The history of singlet formations over $B$ is required to get the
correct coefficient of  $\ln L$ in Eq. (\ref{eq3.3}). From
Eq. (\ref{eq3.3}) we see that singlets form at a constant rate with
respect to an 'RG time' $\ell=\ln \G$. 
Quite generally, we can define the average RG time between Ma-Dasgupta
decimations as
$\overline{\ell}=\overline{\ln \Gamma_1-\ln \G_0}$, where $\Gamma_0$
and $\Gamma_1$ are the RG scales at consecutive singlet formations. We
can write the following general formula:
\be
S_{L} \sim \frac{1}{3}c_{eff}\log_2 L=2\frac{\ln \G_L}{\overline{\ell}}\S_{total},
\label{Sintro}
\ee
where $\ln \G_L$ is the total 'RG time' for a segment of length $L$,
and $S_{total}$ is the average entanglement per Ma-Dasgupta
decimation, which in the Heisenberg model is simply singlet formation,
$\S_{total}=1$.

Next, we find $\overline{\ell}$ in the random singlet phase. Right
after a Ma-Dasgupta decimation step occurs, the resulting bond is strongly suppressed
(see Eq. (\ref{MaDas})); the initial bond strength distribution of
an effective bond forming over a decimated bond at $\G_0$ is
\be
Q(\beta)=\int d\beta_1 d \beta_2
\delta_{(\beta_1+\beta_2-\beta)}P_{\G_0}(\beta_2)P_{\G_0}(\beta_1)=\frac{\beta}{\G_0^2}e^{-\beta/\G}.
\label{eq3.4}
\ee
As the RG progresses, this distribution evolves as $Q_{\G}(\beta)$
with the following evolution equation:
\be
\ba{c}
\frac{dQ_{\G}(\beta)}{d\G}=\frac{\partial Q_{\G}(\beta)}{\partial
  \beta}-2Q_{\G}(\beta)P_{\G}(0)
+2 P_{\G}(0)P_{\G}\stackrel{\beta}\times Q_{\G}.
\label{eq3.5}
\ea
\ee
which is literally what happens when the bond $B$ is next to a
decimated bond. The first term is due to the change in $\beta$ when $\Omega$ changes,
the second and third terms account for $B$'s flow due to one of its two
neighbors forming a singlet. Note that
$\frac{dp_{\G}}{d\G}=-Q_{\G}(0)$. Eq. (\ref{eq3.5}) can be solved
using the ansatz:
\be
Q_{\G}(\beta)=\l(a_{\G}+b_{\G}\frac{\beta}{\G}\rr)P_{\G}(\beta)
\label{eq3.6}
\ee
by substituting Eq. (\ref{eq3.6}) in Eq. (\ref{eq3.5}) we obtain
\be
\ba{cc}
\frac{da_{\G}}{dl}=b_{\G}-2a_{\G}, &
\frac{db_{\G}}{dl}=-b_{\G}+a_{\G}, 
\ea
\label{eq3.7}
\ee
with $l=\ln \G/\G_0$. Also $a_{\G_0}=0,\,b_{\G_0}=1$, from
Eq. (\ref{eq3.4}). 

Now, the probability the the bond $B$ was not decimated again is given by: 
$\int\limits_0^{\infty} d\beta Q_{\G}(\beta)=p_{\G}=a_{\G}+b_{\G}$
and depends on $\G$ {\it only through} $l=\ln \G/\G_0$, reaffirming
our definition of 'RG time' $\ell$. We find:
\be
\overline{\ell}=\int\,dp_{\G}\,\ell=\int\limits_0^{\infty}\,d\ell\,a_{\G} \ell.
\label{eq3.9}
\ee
From Eq. (\ref{eq3.7}) one finds 
\be
\ba{c}
a_{\G}=\frac{1}{\sqrt{5}}\l(e^{-\frac{3-\sqrt{5}}{2}\ell}-e^{-\frac{3+\sqrt{5}}{2}\ell}\rr),\vspace{2mm}\\
b_{\G}=\frac{1}{2}\l[\l(1+\frac{1}{\sqrt{5}}\rr)e^{-\frac{3-\sqrt{5}}{2}\ell}-
 \l(1-\frac{1}{\sqrt{5}}\rr)     e^{-\frac{3+\sqrt{5}}{2}\ell}\rr].
 \label{ab}
 \ea
 \ee
Inserting this in Eq. (\ref{eq3.9}) we find $\overline{\ell}=3$. Therefore:
\be
S_L=\frac{1}{3}\cdot 2 \ln\G+k=\frac{\ln 2}{3}\log_2 L+k.
\label{eq3.11}
\ee
Hence the 'effective central charge' of the random
Heisenberg chain is ${\tilde c}=1\cdot \ln 2$, which is the central
charge of the pure Heisenberg chain times an irrational number: $\ln 2$. 

The same analysis precisely applies to an easy-plane XXZ chain, since
its ground state is also a random singlet fixed point.\cite{RefaelMoore2004} This model also
has $c=1$ in the pure case, and a reduced effective central charge,
$c_{eff}=1\cdot \ln 2$, in the random case. 

\subsection{General formula for random singlet phases}

Now that we have calculated the entanglement entropy of the spin-1/2
random singlet phase, we can easily calculate the entanglement of any
random singlet phase, so long as we know what is the von-Neumann
entropy encapsulated in each site in the system. 

In a random singlet
phase of a system of sites with a
local Hilbert space dimension $D$, we would then have \cite{YangBonesteel}:
\be
S_{L}=\frac{1}{3} \ln L \log_2 D=\frac{1}{3} \ln D \log_2 L.
\label{RSgen}
\ee
For example, the bipartite entanglement in a spin-S random singlet
phase (where strong bonds put pairs of sites into the zero total spin
state) is $S_L=\frac{1}{3}\ln (2S+1)\log_2 L$ \cite{RefaelMoore2}. 

A sufficient condition for the entanglement per singlet to be $\log_2
D$ is that the local Hilbert space is a $D$-dimensional irreducible representation of
some symmetry group, and that the singlet that  two sites form is the
zero-dimensional representation of the group. Under these conditions,
when we look at the density matrix of such two sites, and trace over
one of them, the reduced density matrix, $\hat{\rho}_{red}$ is a $D$
dimensional matrix, which must commute with all group elements,
$\hat{g}\hat{\rho}_{red}=\hat{\rho}_{red}\hat{g}$, and therefore by
Schor's lemma it must be:
\be
\hat{\rho}_{red}=I/D
\ee
with $I$ being the identity matrix. The von Neumann entropy is then
$-tr\hat{\rho}_{red}\log_2\hat{\rho}_{red}=\log_2 D$.
The above arguments would apply also to random $SU(N)$
antiferromagnetic spin-chains, investigated by Hoyos and Miranda
\cite{HoyosMirandaSU3}.

\subsection{Non-abelian random singlet phases}

In Sec. \ref{non-abelian} we reviewed the unique random singlet phases
of non-abelian anyons. The entanglement entropy of these
phases follows the general formula (\ref{RSgen}), but with the
dimension $D$ now being the {\it quantum dimension} of each site. A
rigorous way to prove this is given in
Refs. \cite{YangBonesteel, FRBM}. A simple way of anticipating
the answer, however, is to notice that the Hilbert space of $N$ sites,
each containing a non-abelian charge $a$ with quantum dimension $D$,
will have a Hilbert space of dimension:
\be
dim{\cal H}^N \rightarrow D^N
\ee
as $N\rightarrow \infty$. Since the entanglement entropy of a segment
$L$ in a chain is accumulated from many singlets forming over the
partition bond, connecting the inside of the segment with its
complement, the entanglement per singlet is essentially the log of
the Hilbert space of all the sites on one side of the partition. 

Specific examples are the entanglement of the Majorana chain's random
singlet phase of
Ref. \cite{YangBonesteel}, where $D=\sqrt{2}$, and the bipartite
entropy is:
\be
S_{Majorana}=\frac{1}{3}\cdot\frac{1}{2}\ln 2 \log_2 L
\ee
and the Fibonacci anyon random singlet case, where the quantum dimension $D$ is the
golden mean, $\tau=\frac{1}{2}(1+\sqrt{5})$, with entanglement:\cite{YangBonesteel}
\be
S_{Fibonacci}=\frac{1}{3}\ln \tau \log_2 L=\frac{1}{3}
(0.481211\ldots)\log_2 L.
\label{fiboRSent}
\ee

\subsection{Transverse field Ising model}

The critical point of the random TFIM has the same universal
infinite-randomness scaling as the Heisenberg model. The distributions
of the Ising couplings and the transverse fields are the same. When
considering the entanglement entropy between two segments connected by
the bond $B$, however, if the Ising coupling at $B$ is decimated,
the boundary of the segment becomes a site, with a transverse field
distribution given by Eq. (\ref{eq3.6}) and (\ref{ab}). Entanglement
only occurs once the site straddling the segment boundary is
decimated. After the decimation, the part of the cluster to the right
of the partition becomes a mixed state, and contributes
$\S_{h-dec}=\log_2 2=1$ entropy. Following the boundary-{\it site}
decimation the process repeats, and entanglement accumulates

Thus instead of obtaining entropy $1$ per Ma-Dasgupta decimation, we
only obtain entropy every second Ma-Dasgupta decimation. Referring
back to our general formula for entanglement entropy,
Eq. (\ref{Sintro}), we have $\overline{\ell}=3$, as before, but the
average entanglement per Ma-Dasgupta decimation is $\S_{total}=1/2$. Thus:
\be
S_L=\frac{1}{6}\ln L+k=\frac{\ln 2}{6}\log_2 L+k,
\label{QIE}
\ee
with $k$ a non-universal constant.
The effective
central charge of the random quantum Ising model is ${\tilde
  c}=1/2\cdot \ln 2$ - $\ln 2$ times the central charge of the pure
system. 

\subsection{Generalized TFIM and pure-random c increase}

The xxz, Heisenberg and TFIM all seem to have their effective
central charge reduce as they flow from the pure CFT fixed point, to
the infinite randomness fixed point. Santachiara, however,
demonstrated that this is not the rule using the generalized TFIM
described in Sec. \ref{GTFIM} above. 

The calculation of the entanglement entropy for the random GTFIM is
identical to the that of the random TFIM, with one exception: the
state of the site straddling the partition is, quite generally,
\be
\ket{B}=\summ_{q=1}^{N} \ket{q}_L\ket{q}_R. 
\ee
Therefore upon decimation, the entanglement that results is $\log_2
N$. Since this entanglement is added only every site-decimation, and
not bond decimation, the average entanglement per Ma-Dasgupta
decimation is $\S_{total}=\frac{1}{2}\log_2 N$, and the total
entanglement is:
\be
S_L=\frac{1}{6}\log_2 N\ln L+k=\frac{\ln N}{6}\log_2 L+k,
\label{santa}
\ee
i.e., it has:
\be
c_{eff}^{(N)}=\ln\sqrt{N}.
\ee
When $N\ge 42$, $c_{eff}^{(N)}>c_N$, which means that as the pure
parafermionic $Z_N$ CFT flows to the infinite randomness fixed point,
its entanglement entropy increases, ruling out a pure-random c-theorem. 

\subsection{Entanglement entropy of infinite randomness fixed points
  beyond random singlet.}

As explained in the introduction, there is a hierarchy of infinite-randomness
fixed points with the Damle-Huse exponents $\psi=1/(2s+1)$ and
$\chi=2s$. The calculation of entanglement entropy in non-random
singlet fixed points requires an understanding of the complicated
structures that arise in the quantum state of such points, and the
various Ma-Dasgupta decimations that are possible. For
instance, in the case of a spin-1 permutation-symmetric fixed point
(Sec. \ref{domainwall}), Ma-Dasgupta decimation of a (2,0) domain lying
between two (1,1) domains is different than Ma-Dasgupta decimation
of a (1,1) domain between two (0,2) domains. Similarly, the domain walls
have an internal structure which arises from non-Ma-Dasgupta
decimations, e.g., a ferromagnetic decimation of a (1,1) domain lying
between a (0,2) domain on one side, and a (2,0) domain on the other. 

The strategy for addressing the more complex history dependence is to
revisit the formulation of the entanglement formula,
Eq. (\ref{Sintro}), but allowing for different Ma-Dasgupta decimation configurations:
\be
S_{L} \sim \frac{1}{3}c_{eff}\log_2 L=2\frac{\ln L^{\psi}}{\summ_c p_c\overline{\ell_c}}\summ_c(p_c\S_c).
\label{Sgen}
\ee
The sum over $c$ goes over all possible configurations of the quantum
state at the partition bond at a Ma-Dasgupta
decimation. $\overline{\ell_c}$ is the average RG time for the
configuration $c$ to form, and $\S_c$ is the entanglement entropy of
the configuration.

Without going into details, let us review results for the spin-1 phase
diagram, and for the Fibonnaci anyon phase diagram. 

\subsubsection{Spin-1 entanglement entropy.}

\begin{figure*}\begin{center}
\includegraphics[width=10cm]{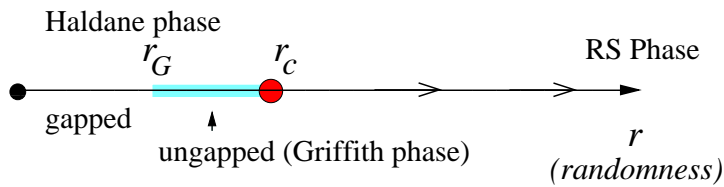}
\caption{Phase diagram of the spin-1 random Heisenberg model. At
  no disorder, $r=0$,
  the chain is in the gapped Haldane phase and its ground
  state resembles a valence-bond solid (VBS). As randomness is
  increased, the gap is destroyed at $r_G$, but the VBS structure
  survives up to the critical point, $r=r_c$. At $r>r_c$ the chain is
  in the spin-1 random-singlet phase. At the critical point, the
  spin-1 permutation symmetric Damle-Huse fixed point obtains, which has
  $\psi=1/2$, and $\chi=2$ \cite{DamleHuse,MGJ,HymanYang}. The
  Entanglement entropy calculation finds that at the permutation
  symmetric point $c_{eff}^{(r_c)}=1.232$, and at the random singlet
  fixed point $c_{eff}^{s=1\,RS}=\ln 3=1.099$. 
\label{spin1PD}}
\end{center}\end{figure*}

The phase diagram of the spin-1 Heisenberg chain is given in
Fig. \ref{spin1PD}. In Ref. \cite{RefaelMoore2004} we calculated the entanglement
entropy of the permutation symmetric fixed point using the method
described above. We found that the leading contribution to the entanglement entropy of the spin-1
random Heisenberg model at the Haldane-RS critical point is:
\be
S\sim\frac{1}{3}{c_{eff}}\log_2 L=\frac{1}{3}\frac{4}{3}\cdot
(1.3327-10^{-3})\cdot \ln 2 \log_2 L.
\ee
where the subtraction indicates the uncertainty in the results, which
is an upper bound. The effective central-charge we find is thus:
\be
c_{eff}^{(r_c)}=1.232
\ee
This effective central charge is smaller than that of the pure system
at the corresponding critical point, $c_{eff}^{r_c}<3/2$. This
effective central charge is also bigger than the effective
central-charges of both the Haldane phase, which vanishes, and the
spin-1 RS phase, which has:
\be
c_{eff}^{s=1\,RS}=\ln 3=1.099.
\ee
Thus in the case of the spin-1 chain, the effective central charge
drops both along flow lines between the pure and random fixed points,
as well as between different infinite randomness fixed points.

\subsubsection{Fibonacci chains entanglement entropy}

The random Fibonacci chain's phase diagram is split between two
infinite randomness phases. The  AFM Heisenberg model flows to the
random-singlet phase, while any density of 'ferromagnetic
couplings', which prefer nearest-neighbors fusing in the $\tau$
channel, destabilizes the random singlet phase, and makes the
chain flow to a mixed infinite-randomness phase, which is in the same
universality class as the Damle-Huse permutation symmetric fixed point for spin-1.

The entanglement entropy of a segment of length $L$ in the random
singlet phase of the Fibonacci chain is easily found in
Eq. (\ref{fiboRSent}); the effective central charge in this phase is
$c_{eff}^{RS}=\ln\frac{1+\sqrt{5}}{2}=0.481211\ldots$. By summing up the contributions of various fusion configurations
between Ma-Dasgupta decimations, and the RG time they require to form,
we find the entanglement entropy of a chain segment with length $L$ in
the mixed FM/AFM phase to be: \cite{FRBM}
\be
S\sim\frac{1}{3}{c_{eff}}\log_2 L=\frac{1}{3}0.702 \log_2 L,
\ee
i.e., the effective central charge is $c_{eff}^{mixed}=0.702$.

It is hard to provide intuition for this result. Nevertheless, it is
interesting to compare it with 
its pure-system analog, and to the effective central charge of the
random singlet phase. It is most likely that the mixed IR phase is
also the terminus of flow from the ferromagnetic pure Fibonacci
chain. The central charge of the critical FM golden chains was
determined in Ref. \cite{Qpeople1} to be
$c=4/5=0.8>c_{eff}^{mixed}$. Hence the effective central charged
dropped along the flow. Comparing our result, though, to the central
charge in the random singlet phase immediately reveals that the
effective central charge {\it increased} in the strong-randomness RG
flow from the random singlet phase to the mixed IR phase. Thus the
suggestion that strong-randomness flows may have a c-theorem
associated with them is contradicted. 

\begin{figure}\begin{center}
\includegraphics[width=8cm]{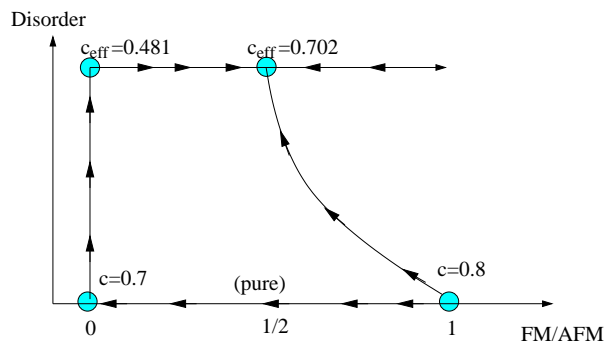}
\caption{Flow diagram of the pure and disordered golden chain. In the pure chain, assuming no intervening fixed points exist, the FM fixed point is unstable to flow to the AFM fixed point, as inferred from the Zamolodchikov c-theorem. In the disordered chain, however, the flow is in the opposite direction, with the mixed  FM/AFM phase, which is most likely the terminus of the flow from the pure FM phase, being stable relative to the random singlet phase,   which is the result of disordering the pure AFM phase. The fixed   point (effective) central charges are also quoted. \label{fullflow}}
\end{center}\end{figure}

Note that by considering the central charges of the of the two
critical phases of the pure chain, $c^{AFM}=0.7$, and $c^{FM}=0.8$, we
find that by the Zamolodchikov c-theorem,\cite{zamolodchikov} the AFM
phase must be stable against FM bond introduction, unless another critical point
appears in between, which we speculate is unlikely. The flow in the
random Fibonacci chain, however, is the opposite: the
mixed FM/AFM phase is stable for essentially all chain coupling
distribution, except for the solely antiferromagnetic point. This
situation and our results are summarized in
Fig. \ref{fullflow}.

\section{Towards measurements of entanglement entropy}

\subsection{Measurement entropy and bounds on entanglement entropy}
The study of entanglement entropy of interacting systems suffers from
the fact that the it is an abstract quantity which is not easy to
measure. Indication that entanglement exists could be found by
confirming Bell inequalities \cite{Bell, Aspect}. Formally,
a theory of entanglement witnesses has been developed
\cite{Horodecki, Terhal, Radim}, but these seem still specific to two
particle systems, and does not provide a quantitative measure of
entanglement.

More recently, the connection of bipartite entanglement and
measurement noise was explored
as a way of quantifying entanglement entropy \cite{KlichRefaelSilva,
  KlichLevitov1,KlichLevitov2}. We define "measurement entropy" of an
observable $\hat{O}$, as the Shannon entropy $S[\hat{O}]=-\sum_{x} P(x) \log
P(x)$ associated with the probability distribution $P(x)$ of the outcomes $x$ of
$\hat{O}$ \cite{Balian}. In classical systems this quantity,
measurable by definition, 
is always lower then the overall entropy. However, in quantum systems it
can in general be either larger or smaller then the entanglement entropy \footnote{Nevertheless
even for observables for which the measurement entropy is larger then the Von Neuman
entropy, information can still be gained in the measurement
\cite{Balian}}. If we consider measurements of local operators as
described below,
we can prove that their measurement entropy provides a lower bound on
entanglement. 

Given a state $ \psi $ of interest, we denote by $\cal L$ the set of observables
$\hat{O}=\hat{O}_A\otimes I+I\otimes \hat{O}_B$, acting locally on $A$ and $B$, for
which $\psi$ is an eigenstate. Let us write the Schmidt decomposition of $\psi$ as $\psi=\sum
c^{\alpha}_{i}|\alpha,i>\otimes|s-\alpha,i>$, where $s$ is the eigenvalue of $\hat{O}$
acting on $\psi$, and such that $\hat{O}_A|\alpha,i>=\alpha|\alpha,i>$ and
$\hat{O}_B|s-\alpha,i>=(s-\alpha)|s-\alpha,i>$ (here $i$ ranges over the degeneracy of
eigenstates of $\hat{O}_A$ with value $\alpha$). The reduced density
matrix can now be written as $\rho_A=\tr_B \rho=\sum
P_{\alpha}\rho_{\alpha}$, where we have defined $\rho_{\alpha}={1\over P_{\alpha}}\sum
|c^{\alpha}_{i}|^2|\alpha,i><\alpha,i| $ and $P_{\alpha}=\sum_i
|c^{\alpha}_{i}|^2$ is the measurement outcome distribution. For the entanglement entropy we have in this case:
\begin{eqnarray}
{\cal S}_{E} = S[\hat{O}_A]-\sum
P_{\alpha}\tr\rho_{\alpha}\log{\rho_{\alpha}}\geq S[\hat{O}_A].
\label{measure}
\end{eqnarray}
where $S[\hat{O}_A]$ is the measurement entropy associated to the probability
distribution $P_{\alpha}$.  This inequality in Eq. (\ref{measure}) is
  completely general. Interestingly, the equality ${\cal
    S}_{E}=S[\hat{O}_A]$ is realized either if all $\alpha$ outcomes are
  non-degenerate, or when the $\rho_\alpha$'s describe pure
  states. The bound (\ref{measure}) becomes better and better by
  choosing a set of commuting operators $\hat{O}$ such that all the
  degeneracy in the measurement result $\alpha$ is removed.

"Conserved" operators are natural candidates for the local operators
  we denote ${\cal L}$ above, i.e. sums of local operators which commute
with the Hamiltonian of the system. For
instance, consider the total spin operator for spin chains with rotational
symmetry.  Generally, the best choice of $\hat{O}$ requires a more elaborate analysis.

\subsection{Entanglement measurement in the random singlet phase}

Bipartite entanglement and fluctuations of a conserved
quantity have a particularly close relationship in the random singlet
phase of the spin-1/2 easy-plane xxz and  Heisenberg chains
The Hamiltonian (\ref{xxz}) only commutes with  $\hat{S}^z_{total}=\summ_i\hat{S}^z_i$
(note, however, that its ground state has a  full rotational symmetry). Therefore
$\hat{S}_A=\summ_{i\in A}\hat{S}^z_{i}$ is the operator of choice for estimating the
entanglement between part A and the rest of the chain. In the random singlet phase
there are two types of singlets: (a) $N_{AB}$ singlets connecting between A and B, (b)
$N_{AA}+N_{BB}$ singlets connecting sites in A to other sites in A, or sites in B to
other sites in B. As explained in Sec. \ref{RSent}, each singlet contributes 1 to the
entanglement entropy, and therefore: ${\cal S}_E=N_{AB}$. 
But in addition, each singlet contributes $1/4$ to the variance of
the measurement of $\hat{S}_A$. Therefore, the random-singlet phase
entropy not only obeys Eq. (\ref{measure}), but can be completely
measured by the relation: 
\be 
{\cal S}_{E}=4 \langle \Delta (\hat{S}^z_A)^2\rangle=N_{AB}. 
\label{var} 
\ee

One possible design for an entanglement measurement through the
variance of $\hat{S}^z_A$ is to use a SQUID with a well defined
flux-pickup region, interacting with a Heisenberg chain realized using
solid-state spins (see, e.g., \cite{Martinis,Ioffe}). A SQUID
that couples, and can measure, the magnetic flux of
electronic spins in part of the chain, essentially carries out a
measurement of $\hat{S}^z_A$, where $A$ is the region of
measurement. Recent measurements of flux qubits have shown that they
are very sensitive to localized two level systems in their vicinity
\cite{Martinis2, Altshuler}; these two-level systems were conjectured
to be electronic spins in Ref. \cite{deSousa:2007p1271,Ioffe}.

\section{Anderson localization}
\label{Anderson}

\subsection{Localization and multifractality}

When quantum particles encounter a spatially random potential, their
wave function may become localized; this is the well known phenomenon of
Anderson localization \cite{anderson59}. To be specific, consider the
following random on-site potential problem:
\be
\H=-\summ_{\langle ij\rangle} t c^{\dagger}_ic_j +\summ_i
V_ic^{\dagger}_i c_i
\label{AH}
\ee
with the on-site potential randomly distributed in the range $-W/2\le
V_i\ le W/2$. At high values of the disorder width, $W/t$, all
eigenfunctions of (\ref{AH}) are localized. Following~\cite{MacKinnon:1981p1199}, in 3d, when $W/t$ reaches the
critical value $w_c=16.3$, delocalized states begin to appear in the
middle of the band. As $W/t$ decreases further, the mobility edge
between localized and delocalized states moves away from the center of
the band, and eventually reaches the bottom (and top) of the
band so that all states are localized.

Apart from major implications for transport properties, the Anderson
localization transition (as a function of single-particle energy, i.e., chemical potential) is associated with universal behavior reflected in the wave function properties. In particular, a localized wave
function at energy $E<E_c$, where $E_c<0$ is the energy at the mobility
threshold, has a localization length which behaves as:
\be
\xi\sim \frac{1}{|E-E_c|^{\nu}}.
\ee
The wavefunction actually exhibits an even richer universal behavior
which is expressed by its {\it multifractality}. Consider the integral of the
wavefunction raised to some power $2q$ over all space. This has a
nontrivial universal dependence on the system size:
\be
P_q(E)=\summ_i \overline{|\psi_i|^{2q}}\sim \frac{1}{L^{\tau_q}} {\cal F}_q\l(L^{1/\nu}(E_c-E)\rr).
\ee
This form is quite general for Anderson localized systems, with
$\tau_q$ and $\nu$ depending on the universality class of the problem
\cite{Evers:2008p1303}. 
For the model Eq. (\ref{AH}), in three dimensions, it was found that $\nu=1.57\pm0.03$
\cite{Slevin:2001p1245}. In the case of a pure system, we have
$|\psi_i|^2q=L^{d(\cdot q-1)}$, where $d$ is the dimensionality of the
system.

\subsection{Single site entanglement entropy and localization}

One measure of entanglement entropy of a wave function is the average
single-site entanglement. This measure can be used to characterize
single-particle wave functions, and it differs from the many-body
bipartite entanglement entropy used above. Its usefulness is also in its
universal properties at critical points~\cite{Kopp:2007p1267}. This
is true for Anderson-type localization transitions as well, as was shown
in~\cite{Jia:2008p1153}, which we review briefly here. 

The average single site entanglement is defined as follows. Write the
wave function as:
\be
\ket{\psi}=\summ_i \psi_i \ket{1}_i \prod_{j\neq i}\ket{0}_j
\ee
where $\ket{n}_i$ indicates $n$ particles at site $i$. It is easy to
see that the reduced density matrix, obtained by tracing over all
occupations of sites {\it except} $i$, is
\be
\rho_i^{red}=|\psi_i|^2 \ket{1}_i {}_i\bra{1} +
(1-|\psi_i|^2)\ket{0}_i {}_i\bra{0}.
\ee
Thus the entanglement between site $i$ and the rest of the system is:
\be
S_i=-|\psi_i|^2\ln(|\psi_i|^2)-(1-|\psi_i|^2)\ln(1-|\psi_i|^2).
\ee
Anticipating that the wave function is localized over a number of
sites, $\ell\gg 1$, we have $|\psi_i|^2\ll 1$. Therefore we can
neglect the second term above, yielding:
\be
\overline{S}_{ss}=\overline{S_i}\approx-\frac{1}{L^D}\summ_i |\psi_i|^2\ln(|\psi_i|^2).
\ee

The scaling properties of the average single-site entanglement for
wave functions with energy $E$ can be
obtained from the multifractal spectrum above:
\be
\overline{S}_{ss}(E)=\l.\der{P_q(E)}{q}\rr|_{q\rightarrow 1}=\l.\ln L P_q(E)\cdot
\der{\tau_q}{q}\rr|_{q\rightarrow 1}+\l.\frac{1}{L^{\tau_q}}\der{{\cal F}_q}{q}\rr|_{q\rightarrow 1}
\label{sssder}
\ee
At criticality this will reduce to:
\be
\overline{S}_{ss}(E_c)\sim \alpha_1 \ln L.
\ee
A metal, or a pure system, would just have $P_q(E)\sim L^{D(q-1)}$,
and therefore:
\be
\overline{S}_{ss}^{metal}\sim D L^{D(q-1)} \ln L.
\ee
Both limits seem to indicate that the first term in Eq. (\ref{sssder})
dominates. Therefore we conclude that the single site average
entanglement in the vicinity of Anderson transitions is:
\be
\overline{S}_{ss}(E) = \ln L {\cal K}(L^{1/\nu}(E-E_c)).
\ee

\section{Numerical studies}
The strongly random quantum critical points that are the main focus of this review can be studied numerically with high accuracy for certain special cases with a free-particle representation.  For example, the random XX model is equivalent via the Jordan-Wigner transformation to a problem of free fermions with random hoppings; the bipartite nature of the hopping leads to a ``particle-hole'' symmetry in the energy.  Similarly the random transverse-field Ising model has been studied extensively and these numerical studies were strong support for the validity of the real-space renormalization group approach.  In this section we will focus on numerical studies of entanglement entropy in random systems, then discuss in the following section some results on higher dimensions.  (Note that a numerical study of single-electron entanglement near the Anderson localization transition was already mentioned in Section~\ref{Anderson}.)

The theoretical calculation reviewed in Section 3 for the disorder-averaged critical entanglement entropy of the random XX model, which leads to a logarithmic divergence with effective central charge ${\tilde c} = c \ln 2$, was confirmed in a subsequent numerical study~\cite{Laflorencie:2005p1056} on systems of up to 2000 sites.  The difference between the entanglement for pure and random critical points is already clear for systems of a few hundred sites, but the larger systems are necessary to see that the divergence is indeed logarithmic.  A key step in the calculation~\cite{Chung:2001p1165} is that the reduced density matrix is determined by the single-particle correlation matrix
\begin{equation}
\left(\ba{ccc}
\langle c_1^\dagger c_1 \rangle & \langle c_1^\dagger c_2 \rangle&\ldots \\
\langle c_2^\dagger c_1 \rangle & \langle c_2^\dagger c_2 \rangle&\ldots\\ 
\ldots && 
\ea
\right)
\end{equation}
which can be computed directly from the {\it single-particle} wavefunctions.

The conclusion that the counting of valence bonds using real-space renormalization group leads to the exact entanglement entropy for random singlet quantum critical points is at first glance surprising.  For pure quantum critical points, it has been shown analytically and numerically~\cite{Jacobsen:2008p916} that ``valence-bond'' entropy and full entanglement entropy differ, although in those examples both have logarithmic divergences.  The valence bond entanglement at random quantum critical points has been conjectured to appear also a universal prefactor appearing in correlation functions, with supporting numerical evidence~\cite{Hoyos:2007p1155}.  (The ordinary correlation function receives nonuniversal contributions as well, but those authors propose a way to cancel the nonuniversal contributions by interfering sublattices.)

Recent work found an interesting relationship~\cite{RodriguezLaguna:2007p1147} between the entanglement entropy of random {\it quantum} critical points in one dimensions and the computational effort required in studying some classical spin glass models by the ``simulated quantum annealing'' approach.  This result is somewhat similar in spirit to the ``finite-entanglement scaling''~\cite{Tagliacozzo:2008p1188,Pollmann:2008p1189} at pure quantum critical points in one dimension, which is determined by the central charge~\cite{Pollmann:2008p1189} via the ``entanglement spectrum'' (the full set of density matrix eigenvalues~\cite{Li:2008p1185}), which was recently determined for conformally invariant critical points~\cite{Calabrese:2008p1439}.  The entanglement spectrum depends on a single parameter combining $c$ and the correlation length, and determines the full set of Renyi entropies (which are essentially moments of the spectrum).

The entanglement spectrum at a random singlet quantum critical point can be conjectured to be rather simple and qualitatively different from the pure case, even though the entanglement entropy is similar.\footnote{We thank P. Calabrese for conversations regarding this point.}  For $N$ bonds crossing the boundary, there are $2^N$ eigenvalues of value $2^{-N}$, which leads to the disorder-averaged Renyi entropy
\begin{equation}
S_\alpha = {1 \over 1 - \alpha} \log_2 \left(\sum_i \lambda_i^\alpha \right) = {1 \over 1 - \alpha} \log_2 2^{N (1- \alpha)} = \log_2 N.
\end{equation}
(This just reflects the additivity of Renyi entropies for independent processes, where each singlet is an independent process.)  Hence a numerical confirmation that the Renyi entropies are equal to the von Neumann entropy in the XX model would show that the random singlet model predicts not just the entanglement entropy but the universal part of the entanglement spectrum.  The challenge may be to separate out this ``universal part'' from non-universal contributions, which is simple in the case of the entanglement entropy.  It would then be interesting to confirm that the Renyi entropies are {\it not} trivially related to the von Neumann entropy in cases where the entanglement is not just a sum over independent singlet bonds.

\section{Higher dimensions and other future directions \label{higherD}}

While entanglement entropy at higher-dimensional critical points is not understood as completely as in one dimension, there are several important results discussed elsewhere in this issue.  In higher dimensions, many gapless systems show the same ``area law'' ($S \sim L^{d-1}$ for spatial dimensionality $d$)~\cite{Eisert:2008p1171,Hastings:2007p1181} as for gapped phases.  An exception is the entanglement for free fermions with a Fermi surface~\cite{gioev,wolf}, which has an additional logarithmic factor that can be related to the many gapless points on the Fermi surface.  Beyond the area law contribution, interesting and sometimes universal subleading terms can appear at some critical points~\cite{Ryu06,fradkinmoore,Mulligan09,sachdeventangle}.    One new aspect in higher dimensions is that the entanglement can have a nontrivial dependence on the geometry of the partition even for the case where one subsystem is a single contiguous region.


Understanding entanglement in some higher-dimensional critical points is possible through the strong-disorder renormalization group approach.  Like all real-space renormalization approaches, dimensions greater than one are considerably more challenging, and some form of numerical analysis of the RG equations is typically necessary.  Two studies to date of the two-dimensional random transverse-field Ising model reached different conclusions regarding entanglement at the model's critical point.  Lin, Igloi, and Rieger found a behavior of the entanglement consistent with an area law times a double logarithm,
\cite{Lin:2007p1042}
\begin{equation}
S(L) \approx L \log_2 \log_2 L + \ldots,
\end{equation}
and gave an argument for this result based on a picture of clusters in the strong-disorder renormalization group.  (For a different model with a close connection to percolation, the bond-diluted transverse Ising model, these authors found an area law in all dimensions.)
Yu, Saleur and Haas also studied the transverse-field Ising model RG equations numerically and found a different result: an area law plus a subleading single logarithm in the disorder-averaged entanglement entropy~\cite{Yu:2008p1046},
\begin{equation}
S(L) \approx \alpha L + \beta \log L + \ldots,
\end{equation}
with $\beta \approx -0.08 \pm 0.01$ and $\alpha$ nonuniversal.  While this is superficially similar to the behavior at certain 2D quantum critical points~\cite{fradkinmoore}, its physical origin is almost certainly different.  The properties of the system were interpreted in terms of percolation behavior, and properties of critical percolation clusters were shown analytically to give a logarithmic term of this form.

While a consensus is yet to emerge, these results do indicate that higher-dimensional critical entanglement entropies are an important remaining challenge: further aspects such as geometry dependence and the connection to other types of random critical points have yet to be studied.  Other major challenges include a better understanding of the (classical and quantum) computational difficulty of random quantum systems and how this connects to entanglement, an understanding of dynamical and thermal properties of entanglement, and a more complete picture of how entanglement is manifested in experimental observations.

We hope that this review has conveyed some of the excitement regarding
the new perspective that entanglement entropy offers on
disordered quantum systems.  The authors wish to thank many collaborators and
colleagues for invaluable conversations over the past five years, and
especially thank N. Bonesteel, P. Calabrese, E. Fradkin, L. Fidkowski, I. Klich,
H.-H. Lin, S. Mukerjee, F. Pollmann, A. Silva, P. Titum, A. Turner, and K. Yang for their entanglement with the authors. The authors gratefully acknowledge financial support from the Packard Foundation,
Research Corporation Cottrell award, and the Sloan Foundation, and NSF
grants PHY-0456720 and PHY-0803371 (GR), and NSF DMR-0804413 (JEM).

\bibliographystyle{unsrt_withcaps}
\bibliography{entrev,entspinrefs}


\end{document}